\documentclass[fleqn,usenatbib]{mnras}

\usepackage{newtxtext,newtxmath}

\usepackage[T1]{fontenc}

\DeclareRobustCommand{\VAN}[3]{#2}
\let\VANthebibliography\thebibliography
\def\thebibliography{\DeclareRobustCommand{\VAN}[3]{##3}\VANthebibliography}

\usepackage{graphicx}	
\usepackage{amsmath}	

\defcitealias{2026arXiv260113785W}{W26}
\defcitealias{2019ApJ...874...32R}{R19}
\defcitealias{2011ApJ...740...92G}{G11}
\defcitealias{2025MNRAS.544..975S}{S25}
\defcitealias{2014MNRAS.445.1898C}{C14}

\title[Still non-accelerating]{Still non-accelerating: age-bias correction in supernova cosmology is robust to host-progenitor age mapping}

\author[C. Chung et al.]
{Chul Chung, 
Junhyuk Son, 
Seunghyun Park, 
Suk-Jin Yoon, 
Hyejeon Cho, 
\newauthor
Dongwook Lim, and Young-Wook Lee\thanks{E-mail: ywlee2@yonsei.ac.kr (YWL)}
\\
Department of Astronomy \& Center for Galaxy Evolution Research, Yonsei University, Seoul 03722, Republic of Korea\\
}

\date{Accepted XXX. Received YYY; in original form ZZZ}

\pubyear{\the\year{}}

\begin{document}
\label{firstpage}
\pagerange{\pageref{firstpage}--\pageref{lastpage}}
\maketitle

\begin{abstract}
We re-examine the claim by \citet{2026arXiv260113785W} that progenitor-age bias has a negligible impact on cosmological inferences from Type Ia supernovae (SNe Ia). 
We show that their inferred host-age–Hubble residual (HR) slope is severely underestimated because their combined SN Ia sample spans an unusually wide redshift range ($0.04 < z < 0.42$), over which the mean host age evolves by $\sim$\,3 Gyr. 
As a result, SNe Ia spanning substantial host-age differences are effectively assigned similar HR values prior to regression, artificially flattening the inferred age–HR relation. 
In addition, their application of the Pantheon+ host-mass correction further suppresses the slope, but the underlying dust model is highly incompatible with the measured dust attenuation curves of galaxies. 
We also demonstrate that our age bias correction is robust to uncertainties in host–progenitor age mapping arising from different choices of the SN Ia delay-time distribution. 
The reduced progenitor-age evolution argued by \citet{2026arXiv260113785W} must, by the same logic, be accompanied by a steeper inferred progenitor-age–HR slope. 
When these two effects are consistently combined in computing the redshift-dependent magnitude correction, the final correction, and hence the resulting cosmological impact, remain largely unchanged from \citet{2025MNRAS.544..975S}.
\end{abstract}

\begin{keywords}
cosmology: dark energy -- distance scale -- transients:supernovae -- galaxies:evolution
\end{keywords}



\section{Introduction}

Supernova cosmology relies on the key assumption that ``the calibrating relationships between Type Ia supernova (SN Ia) luminosity and light curve shape must be invariant with progenitor age'' \citep{2019NatAs...3..706J}. 
Therefore, if this assumption is valid, no correlation should be observed between SN Ia magnitude and progenitor age after luminosity calibration. 
Direct tests of this assumption have become possible only relatively recently, as reliable measurements of the mean age of the stellar populations in SN Ia host galaxies (host age), which is the observable quantity most closely linked to progenitor age, have become available. 
Based on very high quality (signal to noise ratio $\sim 175$) spectra for nearby early-type host galaxies, \citet{2020ApJ...889....8K} first reported a hint of a $\sim 3 \sigma$ correlation between host age and Hubble residual (HR), which is used as a measure of relative SN luminosity after standardization. 
Then, using the \citet[][hereafter \citetalias{2019ApJ...874...32R}]{2019ApJ...874...32R} sample and their photometric age measurements, \citet{2020ApJ...903...22L} found a $4.3 \sigma$ correlation for host galaxies spanning all morphological types. 
This result has since been independently confirmed by two third-party studies at the $\sim 5 \sigma$ level \citep{2021MNRAS.503L..33Z, 2023SCPMA..6629511W}. 
More recently, \citet{2025MNRAS.538.3340C} remeasured ages for a larger sample of host galaxies over a broader redshift range, including the \citetalias{2019ApJ...874...32R} sample at low redshift ($z < 0.2$) and the \citet[][hereafter \citetalias{2011ApJ...740...92G}]{2011ApJ...740...92G} sample at relatively high redshift ($z < 0.42$), and confirmed the ubiquitous nature of the age bias at $5.5 \sigma$ for the \citetalias{2019ApJ...874...32R} sample and $4.8 \sigma$ even for the higher-redshift \citetalias{2011ApJ...740...92G} sample (see Park et al. 2026). 
The origin of this strong correlation was clarified by \citet{2022MNRAS.517.2697L} and Park et al. (2026). 
They showed that, contrary to the key assumption of SN cosmology, the light-curve width-luminosity relation (Phillips relation) and the color-luminosity relation, which are the two correction terms in the SN Ia luminosity calibration process, exhibit significant luminosity offsets depending on host age. 
This is reminiscent of \citet{1956PASP...68....5B}'s discovery of two Cepheid period-luminosity relations that depend on population age, which ultimately led to the realization that the Hubble constant originally derived by Hubble had been seriously overestimated. 
The discovery of an analogous effect in the SN Ia luminosity standardization process suggests that the apparent dimming of high-redshift SNe Ia, on which the discovery of cosmic acceleration and the cosmological constant was based, may in fact have been significantly affected by the evolution of progenitor age with redshift and the resulting systematic bias.

\begin{figure*}
	\includegraphics[angle=-90,scale=0.8]{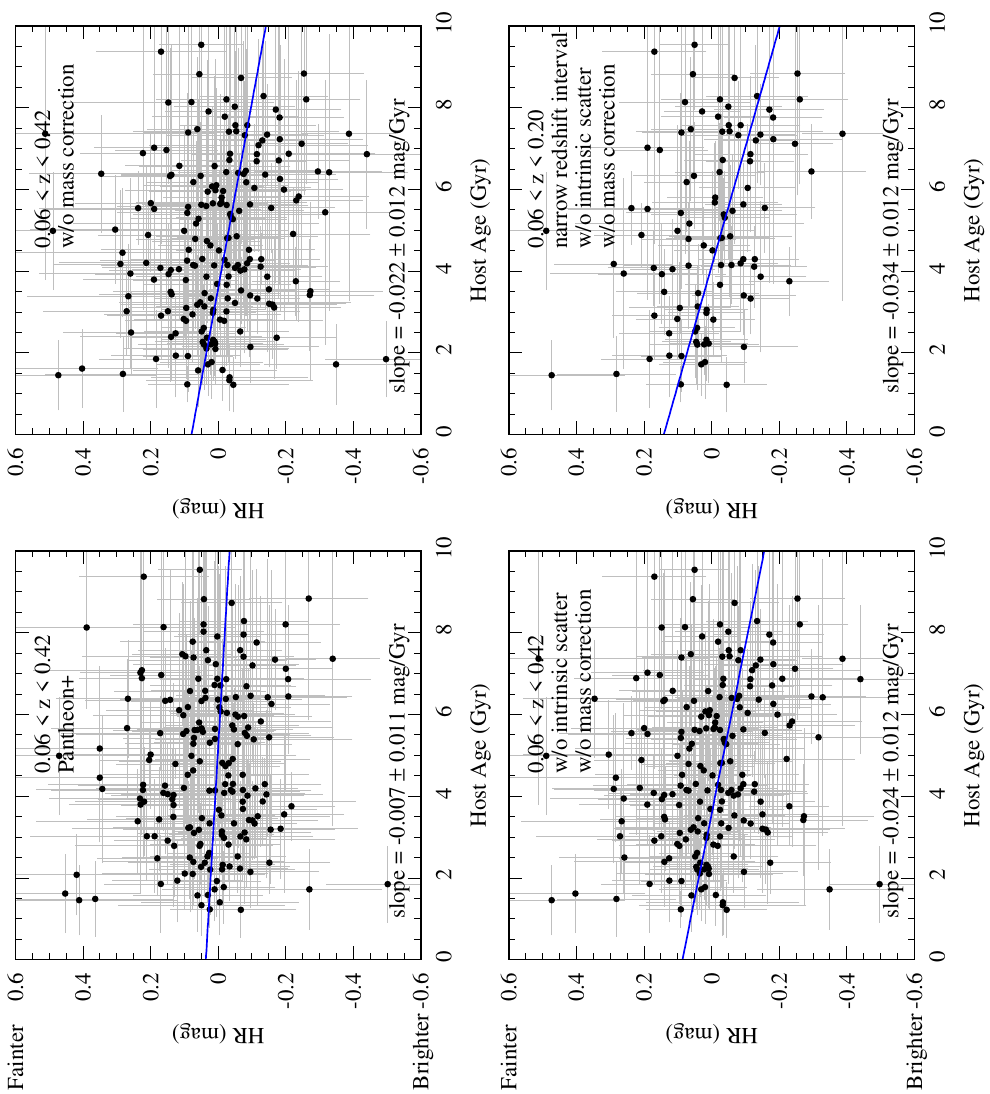}
    \caption{Correlation between host age and HR. 
    As in \citetalias{2026arXiv260113785W}, HRs for the combined \citetalias{2019ApJ...874...32R}+\citetalias{2011ApJ...740...92G} sample are calculated using the Pantheon+ \citep{2022ApJ...938..110B} parameters and methodologies with (upper-left panel) and without (upper-right panel) the host-mass correction. 
    The lower-left panel is obtained when the intrinsic scatter is additionally removed, and the lower-right panel shows the corresponding subsample in the narrow redshift interval $0.06 \leq z \leq 0.20$, which yields the slope closest to the true value (see text).} 
    \label{f01}
\end{figure*}

Based on these results, \citet[][hereafter \citetalias{2025MNRAS.544..975S}]{2025MNRAS.544..975S} showed that, after correcting for the age bias as a function of redshift, the SN dataset becomes more consistent with the $w_0w_a$CDM model recently suggested by the Dark Energy Spectroscopic Instrument (DESI) baryon acoustic oscillation (BAO) project \citep{2025PhRvD.112h3515A} from a combined analysis using only BAO and the cosmic microwave background (CMB) data. 
When the three cosmological probes (SNe, BAO, and CMB) are combined, they found a significantly stronger ($> 9 \sigma$) tension with the $\Lambda$CDM model, suggesting a time-varying dark energy equation of state in a currently non-accelerating universe. 
In their rebuttal, however, \citet[][hereafter \citetalias{2026arXiv260113785W}]{2026arXiv260113785W} argued that the key ingredients underlying this conclusion -- the host-age dependence of standardized SN magnitude, the redshift-dependent progenitor-age difference, and their combined impact -- are either negligible or already corrected for in the Pantheon+ \citep{2022ApJ...938..110B} SN dataset.
In this paper, which serves as a counter-rebuttal to \citetalias{2026arXiv260113785W}, we show that the main arguments advanced by \citetalias{2026arXiv260113785W} are either based on serious methodological flaws or are internally inconsistent by their own logic.

\section{The origin of the shallower slope of the host-age--HR relation in W26}

\citetalias{2026arXiv260113785W} claimed that the slope of the host-age--HR correlation in the combined \citetalias{2019ApJ...874...32R}+\citetalias{2011ApJ...740...92G} sample is substantially weaker, and of low statistical significance, compared to the slopes derived in \citet{2025MNRAS.538.3340C} and \citetalias{2025MNRAS.544..975S} when the \citetalias{2019ApJ...874...32R} and \citetalias{2011ApJ...740...92G} samples are analyzed separately. 
However, their inferred slope is severely underestimated for two reasons. 
First, the redshift range of their combined sample ($0.06 < z < 0.42$) is too wide for HR to be interpreted simply as a measure of relative luminosity. 
HR can be used as a proxy for relative luminosity only when the SNe Ia being compared lie at essentially the same redshift. 
In practice, a narrow redshift bin ($\sim 0.1$) is usually adopted as a compromise to secure a sufficient sample size while still allowing HR to serve this role. 
If the redshift range becomes substantially wider, however, HR is no longer a pure measure of relative luminosity, because its value depends on the baseline cosmological model used in its definition, while the mean age of the host stellar population also evolves with redshift. 
Unlike the \citetalias{2019ApJ...874...32R} sample, which is confined to a relatively narrow redshift range ($0.06 \leq z \leq 0.20$), the combined \citetalias{2019ApJ...874...32R}+\citetalias{2011ApJ...740...92G} sample used by \citetalias{2026arXiv260113785W} spans a much broader interval ($0.06 \leq z \leq 0.42$), over which HR acquires a redshift-dependent contribution from the assumed baseline cosmological model. 
Over this same range, the mean host age evolves by $\sim 3$~Gyr (see Figure 2 of \citetalias{2025MNRAS.544..975S}). 
Because a $\Lambda$CDM model was adopted as the baseline when computing HRs, SNe Ia spanning host-age differences of up to $\sim 3$~Gyr, corresponding to $\sim 40 \%$ of the full age span, were effectively assigned similar HR values even before the host-age-HR regression was performed. 
This procedure inevitably leads to a substantial underestimation of the slope. 
For this reason, in \citet{2025MNRAS.538.3340C}, we analyzed the \citetalias{2019ApJ...874...32R} and \citetalias{2011ApJ...740...92G} samples separately, and for the \citetalias{2011ApJ...740...92G} sample, which spans a broader redshift interval, we further corrected the HR values for redshift evolution.

The second reason is that \citetalias{2026arXiv260113785W} apply a host-mass correction in a context where the goal is to test the relation between host age and standardized SN magnitude.
Because host stellar mass is strongly correlated with host age, applying a mass-dependent correction can partially remove the age-dependent signal itself and thereby bias the inferred age-HR slope toward smaller values. 
This is especially important because the Pantheon+ bias-correction framework incorporates a host-mass-dependent dust total-to-selective extinction ratio, $R_V$, assigning unusually low values of $R_V$ (1.5 or 2.1) to high-mass host galaxies \citep{2021ApJ...909...26B, 2021ApJ...913...49P, 2023ApJ...945...84P}. 
These values differ from the typical Milky Way value, $R_V = 3.1$ \citep{1989IAUS..135P...5C}, and the mean value, $R_V = 3.0$, measured for both early- and late-type high-mass galaxies by \citet[][see their Table 1]{2018ApJ...859...11S}. 
They also lie far outside the range $2.6 < R_V < 3.5$ spanned by the corresponding subsamples of \citet{2018ApJ...859...11S}. 
Thus, this is not merely a modest deviation from the observed values, but a choice that is difficult to justify and is inconsistent with observational constraints. Even more seriously, the required mass-dependent $R_V$ values demand a trend exactly opposite to that indicated by the measured dust attenuation curves of 230,000 galaxies \citep{2018ApJ...859...11S}, and are therefore likewise observationally unsubstantiated. 
Since all galaxies in the \citet{2018ApJ...859...11S} sample would at one time or another have served as Type Ia SN host galaxies, it is difficult to find any reasonable basis for assuming that only the Pantheon+ host galaxies should require such an anomalously low $R_V$ value and a mass dependence opposite to that observed. 
To account for these discrepancies, \citet{2024MNRAS.534.2263P} suggested either that SNe Ia preferentially select the extreme ends of galactic dust distributions, or that the dust properties along SN Ia lines of sight are incompatible with those inferred for galactic dust distributions. 
In the absence of any observational or theoretical basis for such claims, a more natural explanation is that these specific choices most likely arose from somewhat strained attempts to suppress correlations between host properties and HR through a dust model, while disregarding the evidence for the host-age--HR correlation.

Figure~\ref{f01} clearly illustrates these effects step by step. 
Following the \citetalias{2026arXiv260113785W} analysis, we computed HRs and their uncertainties for the combined \citetalias{2019ApJ...874...32R}+\citetalias{2011ApJ...740...92G} sample using the light-curve parameters given in Pantheon+, together with their treatment of covariance, intrinsic scatter, and host-mass dependent dust correction. 
The upper-left panel shows the result obtained when the HRs are calculated directly from the Pantheon+ parameters. 
The resulting slope, $s = -0.007$~mag/Gyr, is nearly zero, consistent with the \citetalias{2026arXiv260113785W} conclusion. 
The upper-right panel shows the effect of removing the host-mass correction, yielding a slope of $s = -0.022$~mag/Gyr, again consistent with \citetalias{2026arXiv260113785W}, but still shallower than that of \citet{2025MNRAS.538.3340C} and \citetalias{2025MNRAS.544..975S}. 
{The Pantheon+ data used by \citetalias{2026arXiv260113785W} provides distance-modulus uncertainties that include the intrinsic error terms introduced to reproduce the unexplained scatter in the Hubble diagram.}
As emphasized by \citetalias{2011ApJ...740...92G}, \citet{2018ApJ...854...24K}, and \citet{2020A&A...644A.176R}, most of this scatter may arise from host-galaxy-dependent effects, so including it in the HR error budget can suppress the very signal being tested and weaken the inferred correlation. 
The lower-left panel shows the result obtained when, in addition to removing the host-mass correction, this intrinsic scatter is also removed from the HR error. 
As expected, the slope increases slightly to $s = -0.024$~mag/Gyr. 
Finally, the lower-right panel shows the result obtained under the same conditions as in the preceding case, but using only the subsample in the narrow redshift interval $0.06 < z < 0.20$, where HR can more reliably serve as a measure of relative luminosity. 
The inferred slope, $s = -0.034$~mag/Gyr, which we regard as the Pantheon+ based estimate closest to the true host-age--HR slope, is substantially steeper than the value reported by \citetalias{2026arXiv260113785W}.

As shown in the upper panels of Figure~\ref{f01}, the host mass-step correction may appear to remove a fraction of the age effect at a given redshift, but it does not correct the age bias relevant for cosmological inference, which depends on how HR varies with redshift. 
The reason is simple: host age and host mass evolve very differently with redshift. 
Over the redshift range most relevant to SN cosmology, $0 < z < 1$, host mass evolution is small or negligible, whereas host age evolves by $\sim 6$~Gyr \citepalias{2025MNRAS.544..975S}. 
Therefore, on average, two host galaxies at different redshifts would have similar masses while having very different progenitor ages, and the resulting age bias would not be corrected by the mass-step correction because the two hosts have nearly the same mass. 
This is a crucial point for cosmological inference, yet it has generally been overlooked in the current SN cosmology community. 
In short, \citetalias{2026arXiv260113785W} used a redshift interval that is far too broad for isolating the age-HR relation at fixed redshift, and further applied a host-mass correction that can also absorb part of the age signal. 
For these reasons, their analysis yields an attenuated estimate of the slope, making it appear much shallower than it actually is.

\section{Host-progenitor age mapping and the age-HR slope}

\begin{figure*} 
    \hspace*{-0.5cm}
	\includegraphics[angle=0,scale=0.40]{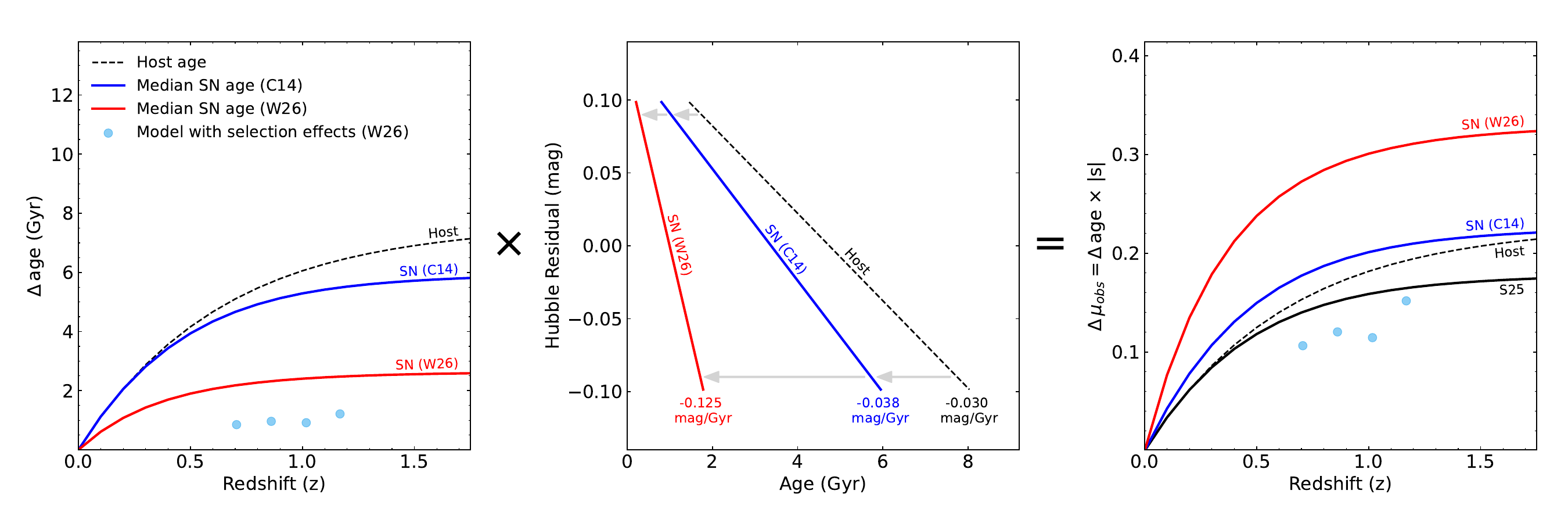}
    \caption{The redshift-dependent magnitude correction (right panel), obtained by combining the median SN progenitor-age evolution (left panel) with the progenitor-age slope (center panel). 
    The blue and red lines correspond to the \citetalias{2014MNRAS.445.1898C} and \citetalias{2026arXiv260113785W} DTDs, respectively, while the cyan circles show the host-galaxy models adopted by \citetalias{2026arXiv260113785W} that further incorporate survey selection effects. 
    The black solid line in the right panel shows the \citetalias{2025MNRAS.544..975S} result for comparison (see text).}
    \label{f02}
\end{figure*}

The second main argument advanced by \citetalias{2026arXiv260113785W} is that, unlike host age, which changes substantially with redshift, SN progenitor age may exhibit a much smaller relative variation over the same redshift interval. 
In \citetalias{2025MNRAS.544..975S}, the expected relative evolution of progenitor age over $0 < z < 1$ was found to be nearly 90\% of that of host age, implying little difference between the two.
This result was obtained using the SN Ia delay-time distribution (DTD) proposed by \citet[][hereafter \citetalias{2014MNRAS.445.1898C}]{2014MNRAS.445.1898C}, which represents an average DTD among those reported in the literature. 
The left panel of Figure~\ref{f02} reproduces this result, showing the redshift evolution of the median SN progenitor age relative to $z = 0.0$, obtained by convolving the cosmic star formation history (CSFH) with the DTD. 
\citetalias{2026arXiv260113785W}, however, argued that adopting a DTD with a much shorter prompt time and a different functional form ({described by index} $\beta$) leads to a significantly weaker redshift evolution of SN progenitor age (see also \citealt{2026arXiv260416597M} for a similar argument). 
When this effect is further combined with their host-galaxy models incorporating survey selection effects, \citetalias{2026arXiv260113785W} argued that the relative difference in progenitor age is reduced to approximately one fourth of that reported by \citetalias{2025MNRAS.544..975S}, and hence that the cosmological impact of progenitor-age bias becomes negligible. 
\citetalias{2026arXiv260113785W} therefore correctly cautioned against the conflation of host age with SN progenitor age and suggested that cosmological analyses should, in principle, be based on progenitor age itself. 
However, their conclusion that the cosmological effect must be negligible is in fact based on an implicit conflation of the two quantities, as described below.

To assess how SN progenitor-age bias affects cosmological inferences, one must evaluate the redshift-dependent magnitude correction, as described in \citetalias{2025MNRAS.544..975S}. 
This correction is obtained by combining the slope of HR with progenitor age ($d{\rm HR}/dt$; hereafter the ``progenitor-age slope'') with the redshift evolution of the median progenitor age relative to its value at $z = 0.0$ (see Figure~\ref{f02}). 
The resulting cosmological impact is therefore determined jointly by the progenitor-age slope and the redshift evolution of progenitor age. 
However, because SN progenitor age cannot be measured directly, the progenitor-age slope must be inferred from the measured host ages and the observed host-age slope using host-galaxy models such as those of \citetalias{2014MNRAS.445.1898C} or {\citet{2022MNRAS.515.4587W}}.
The center panel of Figure~\ref{f02} illustrates how the observed host-age slope ($-0.030$~mag/Gyr; \citetalias{2025MNRAS.544..975S}) is translated into a progenitor-age slope for different choices of the DTD. 
For this purpose, we use the \citetalias{2014MNRAS.445.1898C} host-galaxy models, which provide a relation between host mass and SFH (and hence mass-weighted host age). 
Based on these models, we estimate progenitor ages from the measured host ages and convert the observed host-age slope into a progenitor-age slope over the relevant host-age range $1.5 < t < \sim 8$~Gyr, within which the {\citetalias{2026arXiv260113785W} predicts} a much smaller variation in progenitor age. 
For the \citetalias{2014MNRAS.445.1898C} DTD adopted in \citetalias{2025MNRAS.544..975S}, the difference between host age and progenitor age is relatively small, so the inferred progenitor-age slope is only about 25\% steeper than the host-age slope. 
However, when the same calculation is carried out using the DTD advocated by \citetalias{2026arXiv260113785W}, we find that the progenitor-age slope inferred from the host-age slope becomes approximately four times larger than the host-age slope itself. 
This follows naturally because the denominator entering the progenitor-age slope conversion becomes substantially smaller, leading to a correspondingly steeper inferred slope. 
Therefore, a reduction in the redshift evolution of progenitor age is accompanied by an increase in the progenitor-age slope. 
When these two effects are combined to compute the redshift-dependent magnitude correction (Figure~\ref{f02}, right panel), the final correction becomes nearly twice as large as that of \citetalias{2025MNRAS.544..975S} if only the DTD is changed, and remains comparable to the \citetalias{2025MNRAS.544..975S} result (about 85\%) even when the selection effect advocated by \citetalias{2026arXiv260113785W} is additionally included. 
When SNe, BAO, and CMB are combined in the cosmological analysis, this leads to nearly the same cosmological conclusion as in \citetalias{2025MNRAS.544..975S}.

\section{Discussion}

In this counter-rebuttal to \citetalias{2026arXiv260113785W}, we have shown that the two main arguments advanced in their paper are either affected by serious errors or lead to conclusions that are internally inconsistent by their own logic. 
The first issue concerns their substantially underestimated host-age slope. 
One reason for this is that the redshift interval of their sample is far broader than the bin size normally adopted to test this effect. 
Over such a broad redshift interval, HR is no longer a reliable measure of relative luminosity, causing the host-age slope to be artificially underestimated. 
The other reason is that they further reduce the slope by additionally applying a host-mass correction. 
This is plainly inappropriate, because when one is explicitly testing the host-age effect, applying at the same time a correction based on host mass, which is itself closely tied to age, will inevitably suppress the slope relative to its true value. 
Moreover, the Pantheon+ host-mass correction requires both an anomalously low $R_V$ value for high-mass galaxies and a mass dependence opposite to that indicated by observations. 
The second issue raised by \citetalias{2026arXiv260113785W} concerns the conflation of host age with SN progenitor age. 
Although this caution is valid, they overlook the fact that the same effect that suppresses progenitor-age evolution simultaneously steepens the inferred progenitor-age slope. 
As a result, they incorrectly conclude that the cosmological effect of progenitor-age bias must be negligible.

As discussed above, converting directly measured host ages into progenitor ages may in principle seem necessary for evaluating the cosmological impact. 
In practice, however, the result remains strongly affected by the uncertainty in the still poorly constrained DTD. 
Therefore, an empirical approach based on the directly observed evolution of host age and the measured host-age slope may be practically more reliable than converting host age into progenitor age. 
Historically, many cosmological studies have been conducted, and continue to be conducted, through such empirical approaches. 
For the high-precision SN cosmology studies of the LSST era, however, a more robust cosmological test will be required. 
Such a test should be free from uncertainties in the DTD and the resulting host-progenitor age mapping, from possible systematic bias introduced by arbitrary $R_V$ choices in dust-extinction corrections, from the host-age-HR correlation itself, and from the host-age-dependent zero-point offset in the WLR and the resulting overcorrection in the luminosity standardization process. 
To meet these requirements, we propose an evolution-free test for SN cosmology. 
This approach would use only homogeneous SNe Ia occurring in young and coeval host galaxies over the full redshift range. 
In this way, one could perform the most precise cosmological test with SNe Ia without having to convert host age into progenitor age or apply a redshift-dependent magnitude correction driven by progenitor-age evolution. 
In fact, this approach was already suggested 28 years ago in the original discovery paper \citep{1998AJ....116.1009R}, and it remains the most reasonable and practical path forward for SN cosmology.

\section*{Acknowledgements}

We acknowledge support from the National Research Foundation of Korea to the Center for Galaxy Evolution Research (RS-2022-NR070872, RS-2022-NR070525).
S.-J.Y. acknowledges support from the Mid-career Researcher Program (RS-2024-00344283) through Korea's NRF funded by the Ministry of Science and ICT.

\section*{Data Availability}

There are no new data associated with this article.



\bibliographystyle{mnras}
\bibliography{references} 








\bsp	
\label{lastpage}
\end{document}